\documentclass[%
reprint, 
showkeys,
twocolumn,
superscriptaddress,
 amsmath,amssymb,
 aps, physrev,
floatfix,
]{revtex4-2}

\usepackage{graphicx}
\usepackage{dcolumn}
\usepackage{bm}

\begin{document}

\preprint{APS/123-QED}

\title{\textbf{Maximum limit of connectivity in rectangular superconducting films with an oblique weak link.} 
}%

\author{F. Colauto}
\affiliation{Departamento de F\'{i}sica, Universidade Federal 
de S\~{a}o Carlos, 13565-905, S\~{a}o Carlos, SP, Brazil}
 \email{Contact author: fcolauto@df.ufscar.br}

\author{D. Carmo}
\affiliation{Departamento de F\'{i}sica, Universidade Federal 
de S\~{a}o Carlos, 13565-905, S\~{a}o Carlos, SP, Brazil}
\affiliation{Laborat\'{o}rio Nacional de Luz S\'{i}ncrotron, Centro Nacional de Pesquisa em Energia e Materiais, 13083-100, Campinas, SP, Brazil}

\author{A. M. H. de Andrade}
\affiliation{Instituto de F\'{i}sica, Universidade Federal do 
Rio Grande do Sul, 91501-970, Porto Alegre, RS, Brazil}

\author{A.~A.~ M.~Oliveira}
\affiliation{Instituto Federal de Educa\c{c}\~{a}o, Ci\^{e}ncia e Tecnologia 
de S\~{a}o Paulo, 13565-905, S\~{a}o Carlos, SP, Brazil}

\author{M. Motta}
\affiliation{Departamento de F\'{i}sica, Universidade Federal 
de S\~{a}o Carlos, 13565-905, S\~{a}o Carlos, SP, Brazil}

\author{W. A. Ortiz}
\affiliation{Departamento de F\'{i}sica, Universidade Federal 
de S\~{a}o Carlos, 13565-905, S\~{a}o Carlos, SP, Brazil}

\date{\today}

\begin{abstract}
A method for measuring the electrical connectivity between parts of a rectangular superconductor was developed for weak links making an arbitrary angle with the long side of the sample. The method is based on magneto-optical observation of characteristic lines where the critical current makes discontinuous deviations in the flow direction to adapt to the non-uniform condition created by the presence of the weak link. Assuming the Bean critical state model in the full penetration regime for a sample submitted to a perpendicular magnetic field, the complete flow pattern of screening currents is reconstructed, from which the transparency of the weak link, i.e., the ratio between its critical current and that of the pristine sample, $\tau = \frac{J_i}{J_c}$, is then related to the angle $\theta$ formed by two characteristic discontinuity lines which, in turn, are intimately associated to the presence of the weak link. The streamline distribution is compared with magneto-optical observations of the flux penetration in Nb superconducting films, where a weak link was created using focused ion beam milling. The present work generalizes previous analyses in which the weak link was perpendicular to the long sides of the rectangular sample. Equations and measurements demonstrate that the relationship between the transparency and the angle $\theta$ is not affected by the tilting of the weak link. Noticeably, in order to attain optimum connectivity, the weak link critical current can be less than that of the pristine sample, namely, $\tau _{max}=\sin \Phi$, where $\Phi$ is the tilt angle of the weak link. This expression generalizes the previous result of $\tau _{max}=1$ for $\Phi=$ 90$^\circ$.
\end{abstract}


\maketitle


\section{Introduction}

Weak links (WLs) are inherent to granular superconducting materials~\cite{dersch_new_1988, ishida_fundamental_1990, hilgenkamp2002grain,foltyn2007materials,graser_how_2010}. They can also be formed by growing misoriented bicrystals~\cite{Polyanskii_Magneto_1996,Guth2004,palau2004simultaneous,palau2007simultaneous}, created artificially in films by invasive tools~\cite{dobrovolskiy2012electrical, dobrovolskiy2017abrikosov, johansen_transparency_2019}, or when joining bulk superconductors~\cite{Mikheenko_magneto_2012}. Due to the limited electrical connectivity of a weak link ~\cite{Likharev_superconducting_1979} as compared to that of its neighborhood $-$ usually the pristine material $-$ screening currents arriving at the junction are not fully transmitted through; some part, limited by the intergranular critical current, $J_i$, passes on, while the remaining is forced to recirculate~\cite{Goldfarb1991DoubleTransition}. 

Current flow and connectivity in superconducting samples with WLs are crucial in some of their main applications, which depend on their capability to conduct electric current without any loss~\cite{graser_how_2010,wang_how_2017}, 
a feature which is downgraded by the presence of weak connections. In fact, WLs are present in a vast variety of situations, from polycrystalline samples with grain boundaries~\cite{prester_current_1998} to bulky apparatus based on two or more superconducting pieces connected to form a large-scale device. For instance, high-temperature bulk magnets can trap magnetic fields of several tesla~\cite{tomita_high-temperature_2003}, and by combining such elements, a large assembly can be formed. Superconducting magnets represent another example where at least two superconducting joints are needed to form a persistent current switch. Similarly, mesoscopic tunnel junctions, essential to some superconducting devices such as SQUIDs, as well as high-speed superconducting electronics, and microwave sources~\cite{clarke_squid_2004}, include WLs.

Different approaches can be employed to measure the electrical connectivity between joined superconductors. Regarding magnetic and magneto-transport measurements, granular materials exhibit a double transition when either the temperature, the applied magnetic field, or the transport current is varied. Connectivity can be evaluated as long as the intergranular transition is reasonably narrow ~\cite{Goldfarb1991DoubleTransition,araujo1999multilevel,passos2001granularity,passos2002compiling,ortiz_vortex_2006}. The critical current of a joining interface can be detected $-$ and eventually quantified $-$ by the levitation force when displacing a small magnet over a surface area that includes the WL~\cite{kordyuk_simple_2001}. Alternatively, Hall probes can be used to scan such area~\cite{bending_local_1999}. Some of the methods mentioned above give global averages of the measured property, while others provide a spatial map of the evaluated quantity. When the specimen investigated is a film $-$ or even a bulk parallelepiped with flat enough faces, the most suitable technique to detect and measure the connectivity is magneto-optical imaging (MOI), which provides a real-time map distribution of penetrated magnetic flux, also revealing the overall formation of current domains~\cite{Polyanskii_Magneto_1996, palau2007simultaneous,johansen_transparency_2019,prester_current_1998}. 

In this work, MOI was applied to probe the flux penetration pattern in two-grain rectangular thin superconducting films divided by one perpendicular or oblique WL. The WL consists of a tiny straight groove $-$ where superconductivity is partially suppressed $-$ created by exposition of the pristine material to a focused ion beam (FIB)~\cite{valerio-cuadros_superconducting_2021}. As the screening currents flow around the film, they are forced to change direction to follow the rectangular shape of the sample. The presence of a WL represents another constraint forcing abrupt changes in the flow direction. At these regions, where the screening current streamlines bend, a visible line appears in the flux density distribution, commonly called discontinuity line, or simply d-line~\cite{schuster_discontinuity_1995}. The present study aims to measure the electrical connectivity of the joined interface through the readily observable angle $\theta$ between the d-lines resulting from the presence of the WL. A previous expression for the transparency in a simpler case, where the WL was perpendicular to the longer sides of the rectangle, has now been generalized for any angle $\Phi$ between an oblique WL and the longer sides. Two key results emerge: (i) the relationship between the transparency, $\tau$, and the angle $\theta$, is not affected by the tilting of the WL, that is, $\tau=\cos \theta$ regardless of the angle of inclination; and (ii) in general, the case of optimum connectivity $-$ i.e., maximum transparency $-$ is achieved for values of the WL critical current $J_i$ smaller than that of the pristine sample, $J_c$, namely, $\tau _{max}=\sin \Phi$. This expression generalizes the previous result of $\tau _{max}=1$ for $\Phi=$ 90$^\circ$. 

\section{Materials and Methods}

Nb films were deposited on a Si(100) substrate via magnetron sputtering in an AJA Orion-8 UHV chamber with a base pressure below $2 \times 10^{-8}$ Torr. 
The films analyzed in this study have a nominal thickness of 200 nm and were shaped into rectangular samples, measuring $2.5 \times 0.8$ mm$^{2}$, using optical lithography. 
FIB milling was employed to thin the films along a straight line between the longer sides of the rectangle, creating samples divided by a groove into two homogeneous parts connected by a uniform WL. 
The grooves were formed using a JEOL JIB-4500 Multi Beam System SEM-FIB, set to deliver the minimum Ga$^+$ dose of 0.1~nC/$\mu$m$^2$.

The MOI observations were conducted in a setup where a Bi-substituted ferrite garnet film (FGF) with in-plane magnetization was directly placed on top of the sample, functioning as a Faraday-active flux density sensor.
The FGF was made by liquid epitaxial growth on gadolinium gallium garnet substrates~\cite{helseth_faraday_2002}.
In a polarized light microscope with crossed polarizers, MOI enables real-time observation of magnetic flux penetration into superconducting samples~\cite{vlasko-vlasov_magneto-optical_1999, Polyanskii_magneto-optical_2004}.

The optical cryostat employed in this work is an Oxford MicrostatHe-R, where the sample is mounted on a cold finger at the end of a bayonet.
The sample is inserted into the cryostat beneath an optical window, enabling the MOI to be performed.
Magneto-optical images of the samples were captured using a BX-LRA-2 Olympus microscope equipped with a U-PO3 polarizer, U-AN360-3 analyzer, and MPLFLN 5X objective, while a Retiga 4000R Q-imaging CCD camera recorded the images.
Magnetic fields were applied perpendicularly to the samples using a Helmholtz coil configuration.
Magneto-optical images acquired in such an experimental setup are composed of pixels whose brightness corresponds to the perpendicular magnetic flux density. 
All experiments were conducted at temperatures above 5.0 K to prevent intermittent flux dynamics caused by thermomagnetic instabilities in the Nb samples~\cite{colauto_controlling_2021}.
  
\section{Results and discussion}

\begin{figure}[b!]
\centering\includegraphics[width=1.0\linewidth]{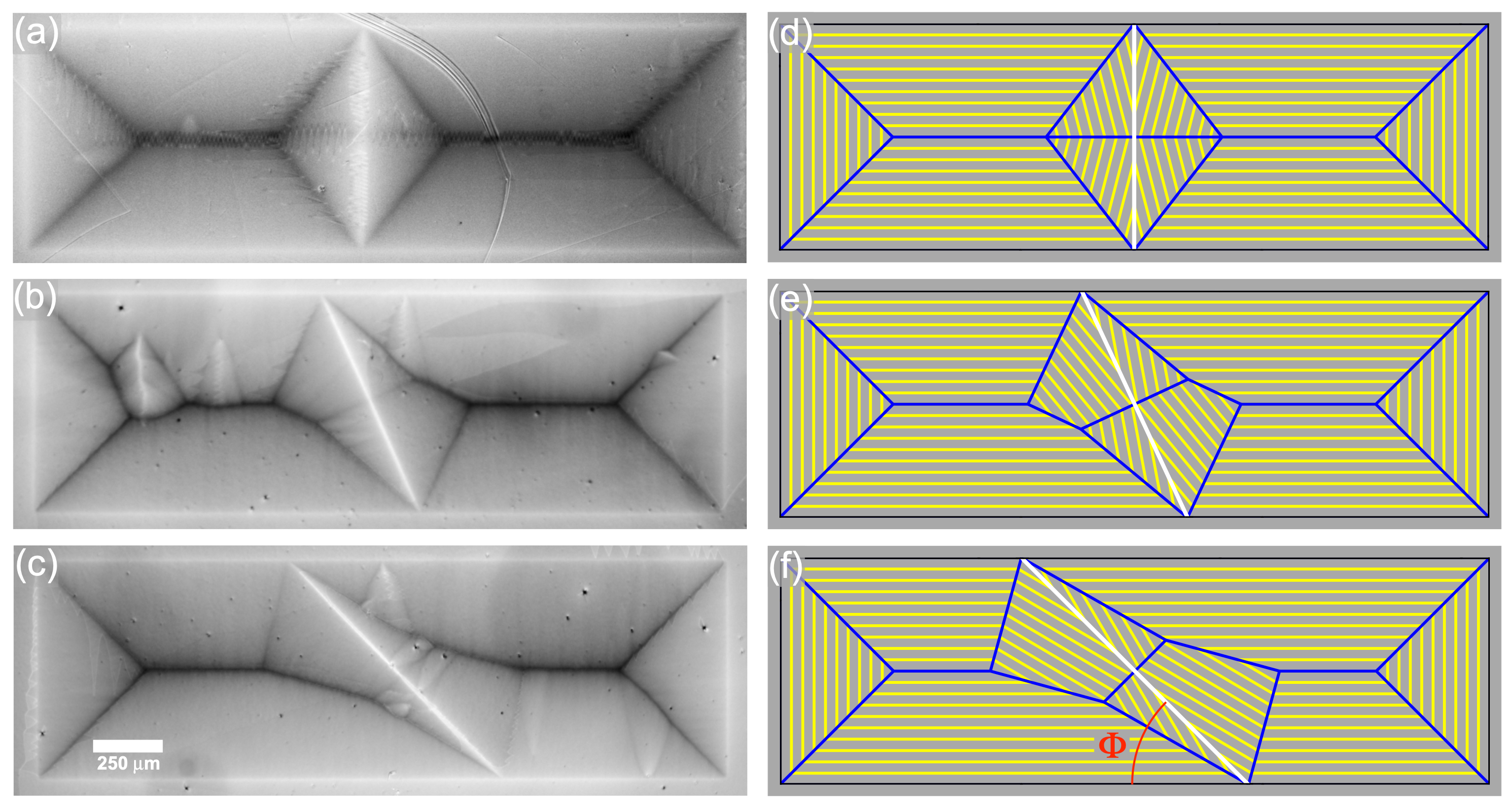}
\caption{
Left column: Magneto-optical images showing flux distribution in Nb films. 
The bright line crossing in the middle is the WL extending between the sample's long edges at angles $\Phi$~= (a)~90$^{\circ}$, (b)~65${^\circ}$, and (c)~45$^\circ$.
D-lines (dim regions) indicate abrupt changes in the direction of the critical current.
Right~column: Reconstruction of critical current streamlines (yellow). 
The WL tilt angles $\Phi$ (white) are: (d)~90$^{\circ}$, (e)~65${^\circ}$, and (f)~45$^\circ$. The d-lines appear in blue. 
The transparency in the patterns is set to $\tau$~=~0.25.
Localized deviations from the expected patterns are due to local defects in the studied film, without any significant distortion in the central d-lines. There are also dark scratches and black spots originating from defects in the MO indicator, which do not affect flux penetration into the sample.
}
\label{Fig1}
\end{figure}

Magneto-optical images displaying flux distribution in samples with WLs are presented in the left column of Figure~\ref{Fig1}.
The angle between the horizontal edge and the WL for each sample is $\Phi=$ 90$^\circ$, 65$^\circ$, and 45$^\circ$ for images (a), (b), and (c), respectively.
Each device was initially zero-field-cooled to 6.0 K, after which a perpendicular magnetic field was gradually increased from zero to 145 Oe, resulting in full flux penetration and inducing screening superconducting currents throughout the specimens.

In the Bean critical state model~\cite{Bean1962,bean_magnetization_1964,Zeldov1994}, the critical current density is $J_c$ throughout the sample, except along the groove where the current density is $J_i$, which is lower than $J_c$ due to the locally reduced thickness, as well as FIB-induced defects $-$ essentially due to Ga$^+$ ions implanted in the remaining material below the groove~\cite{dobrovolskiy2012electrical,chaves_magnetic_2023}. This creates significant disturbances in the current flow around the WL, represented by the white lines in the panels on the right column of Figure~\ref{Fig1}. 
The current streamlines, indicating the flow direction, are shown in yellow for a transparency of $\tau=0.25$. Panels (d) to (f) illustrate the patterns resulting from WLs at angles $\Phi=$ 90$^\circ$, 65$^\circ$, and 45$^\circ$, respectively. 
The figure reveals five distinct current domains on each side of the WLs, representing regions with different unidirectional flows of $J_c$.
At domain boundaries, abrupt changes in flow direction enhance shielding and become visible through magneto-optical imaging, as observed in the panels (a) to (c) of Figure~\ref{Fig1}. 
In these regions, the current lines become discontinuous and are thus referred to as discontinuity lines (d-lines). These d-lines appear dark when the local field is decreasing. In panels (d) to (f), the d-lines are shown as blue segments that emerge spontaneously during the reconstruction of streamline patterns using the Bean model, which simplifies geometric relationships without changing the physics situation. Also evident from Fig. \ref{Fig1} is the fact that all d-lines emanating from the vertices form $45^\circ$ angles with their neighboring edges, as expected for isotropic materials~\cite{brandt1995square,schuster1996flux,colauto_anisotropic_2019}.

Although some phenomena need to be explained using a B-dependent $J_c$~\cite{chaves_magnetic_2023,Jiang2020Selective}, in most cases, the approximation given by the Bean model, in which $J_c(B) = J_c$, is sufficient~\cite{Shantsev2000JcB}. A numerical solution leading to the current streamlines could be obtained through an appropriate procedure to invert the Biot-Savart law~\cite{pashitski_reconstruction_1997}. 
However, our approach clarifies the relationships among all relevant angles related to d-lines, eliminating the need for such computational methods. Since the d-lines are readily observable, attention has turned to the relationship between d-line angles and the transparency of the WL.

\begin{figure}[t!]
\centering\includegraphics[width=1.0\linewidth]{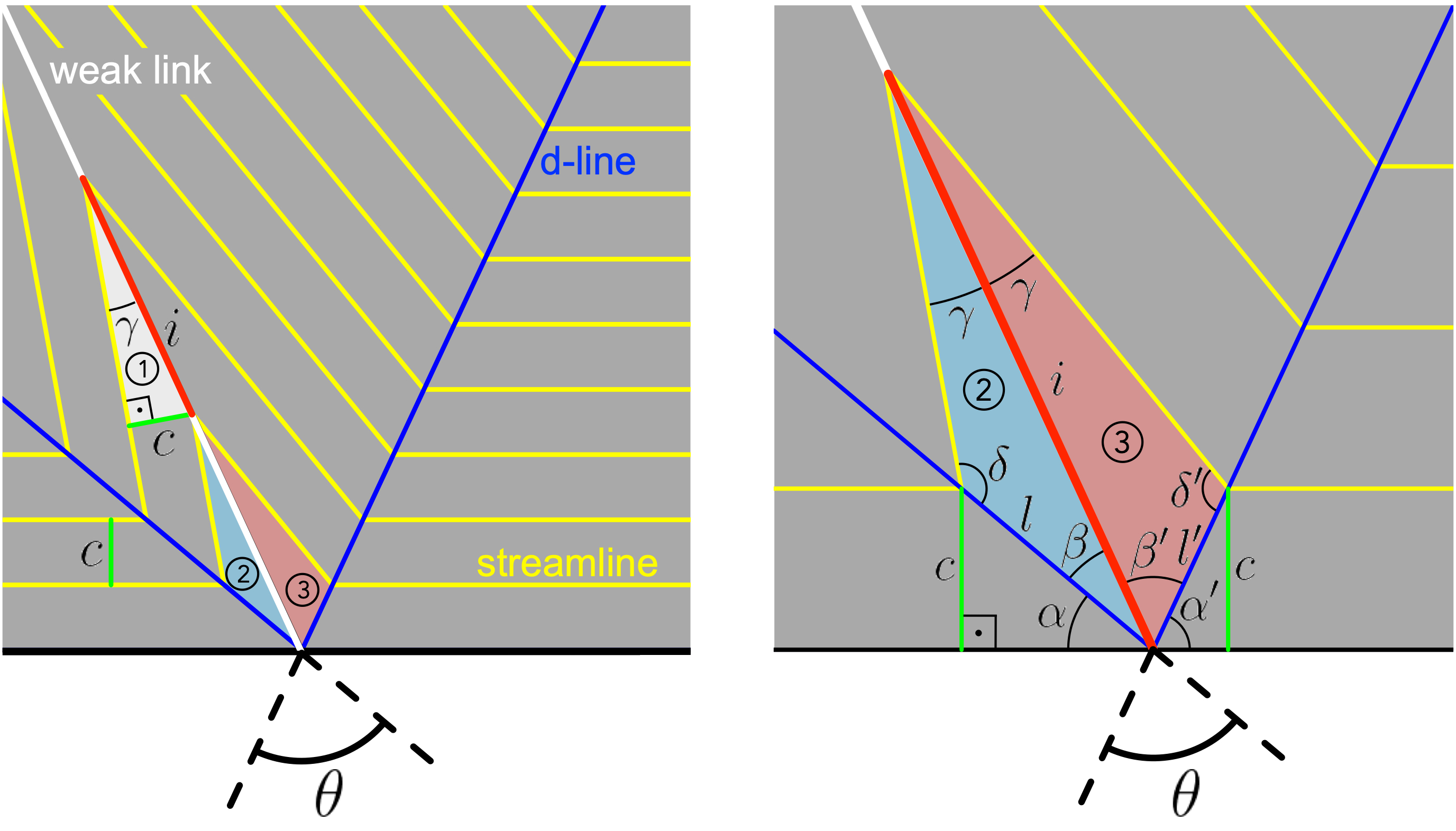}
\caption{Left: Enlargement of critical current streamlines (yellow) around the WL (white), with $c$ and $i$ indicating the distances of current lines at densities $J_c$ and $J_i$, respectively. The transparency is measured by the observable angle $\theta$ between the d-lines (blue). Right: Close-up view highlighting the key angles.
}
\label{Fig2}
\end{figure}

As the critical current is homogeneous in the scope of the Bean model, all streamlines are equivalent, with a constant separation. 
Some important constraints are to be considered in the reconstruction of the streamlines: (i) far from the WL they have to be aligned parallel to the contour of the specimen; (ii) when crossing a d-line, they must form equal angles on both sides, to warrant field-reversal symmetry, meaning that the pattern remains unchanged if the current flow is reversed by applying the magnetic field in the opposite direction; (iii) the distance $c$ between neighboring streamlines represents $\frac{1}{J_c}$ anywhere in the sample except at the WL, where the spacing is $i \ge c$, since $J_i \le J_c$.
As shown in Fig. \ref{Fig2}, triangle 1 (white), the transparency of the WL can be expressed as:
\begin{equation}
    \tau \equiv \frac{J_i}{J_c}= \frac{c}{i} = \sin\gamma.
    \label{Eq01}
\end{equation}
We can apply the law of sines to triangle 2 (blue) on the left side of the WL and find,
\begin{equation}
    \frac{\sin\gamma}{l}= \frac{\sin\delta}{i}.
    \label{Eq02}
\end{equation}
The segment $l$ can be expressed in terms of $c$ and the angle $\alpha$,
\begin{equation}
    l=\frac{c}{\sin\alpha},
    \label{Eq03}
\end{equation}
and the internal angles are related by
\begin{equation}
    \beta + \gamma + \delta = 180^\circ,
    \label{Eq04}
\end{equation}
from which we conclude that 
\begin{equation}
    \alpha = \beta + \gamma.
    \label{Eq05}
\end{equation}

Similarly, on the right side of the weak link, triangle 3 (red) gives us the following equations:
\begin{equation}
    \frac{\sin\gamma}{l'}= \frac{\sin\delta'}{i},
    \label{Eq06}
\end{equation}

\begin{equation}
    l'=\frac{c}{\sin\alpha'},
    \label{Eq07}
\end{equation}
and
\begin{equation}
    \beta' + \gamma + \delta' = 180^\circ.
    \label{Eq08}
\end{equation}
so that 
\begin{equation}
    \alpha' = \beta' + \gamma.
    \label{Eq09}
\end{equation}
Additionally, from the right panel of Fig. 2 one sees that
\begin{equation}
    \alpha + \beta + \alpha' +\beta' = 180^\circ.
    \label{Eq010}
\end{equation}

Substituting Eqs. \ref{Eq05} and \ref{Eq09} into Eq. \ref{Eq010}, and recognizing that $ \theta = \beta + \beta'$, we obtain
\begin{equation}
    \gamma = 90^\circ - \theta.
    \label{Eq011}
\end{equation}

Finally, replacing Eq. \ref{Eq011} into Eq. \ref{Eq01} we obtain the transparency as a function of the observable angle $\theta$.
\begin{equation}
    \tau=\cos\theta.
    \label{Eq12}
\end{equation}

From this result, which holds for any value of the inclination of the WL, one can recover the particular case of $\tau=1$ for $\Phi=90^\circ$ \cite{johansen_transparency_2019}. Equally important, the expression for the transparency is independent of the inclination angle $\Phi$.

\begin{figure}[t!]
\centering\includegraphics[width=1.0\linewidth]{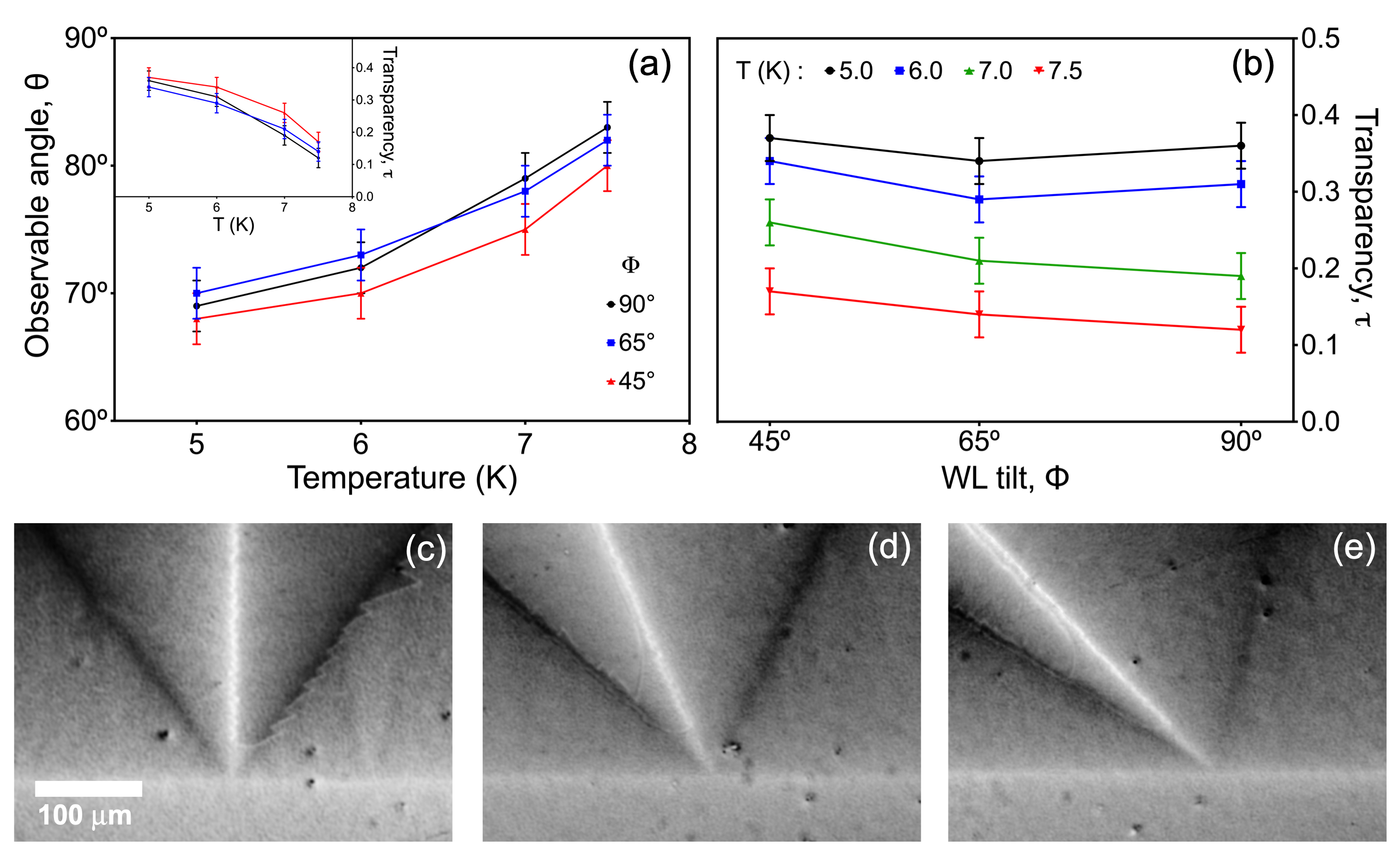}
\caption{Top: (a)~temperature dependence of the observable angle $\theta$ and the transparency $\tau$ (inset). (b)~observable angle versus tilt angle at four different temperatures. Bottom: Magneto-optical images at 6.0~K and 145~Oe for $\Phi$ values of (c)~90$^{\circ}$, (d)~65${^\circ}$, and (e)~45$^\circ$. 
Within the estimated error bar of $\pm~2^\circ$, the observable angle does not depend on the tilt angle of the WL.
}
\label{Fig3}
\end{figure}

Figure 3 shows the observable angle $\theta$ and the transparency as obtained from magneto-optical images. 
Before each measurement, the sample underwent zero-field cooling from a temperature above $T_c$, followed by applying a sufficiently strong perpendicular magnetic field $H^*$ to ensure full penetration.
At $H^*$, all d-lines are visible, and further increasing the field does not affect them as long as $\tau$ remains independent of the field. 
Panel (a) shows that $\theta$ increases around $10^\circ$ from $5.0$~K to $7.5$~K, a trend seen for all WL inclinations studied here.
As the d-lines appear somewhat blurred, a typical error of $\pm~2^\circ$ was estimated for all measurements. 
In view of this uncertainty, the observable angles for various WL inclinations are essentially similar at a given temperature.
The transparency in the inset to panel (a) is derived from Eq. \ref{Eq12}.
Notably, $\tau$ decreases from 0.3 to 0.1 within the measured temperature range, indicating a gradual disconnection of the two grains as the temperature increases. 
Additionally, one sees in panel (b) that transparency is unaffected $-$ within the error bars $-$ by the WL's inclination.

The observable angle can be visually evaluated for samples with different WL inclinations in panels (c) to (e) of Fig. \ref{Fig3}.
These images were taken at 6.0 K and 145 Oe, with the regions of interest enlarged while maintaining proportions.
The bottom horizontal bright lines indicate the edges of the samples, while the vertical or oblique bright lines are the WLs, and the V-shaped dark lines are the d-lines. 
The internal angles of the V-shaped d-lines in these images are $(72\pm2)^\circ$, $(73\pm2)^\circ$, and $(70\pm2)^\circ$, related to $\Phi$ = 90$^\circ$, 65$^\circ$, and 45$^\circ$, respectively.
Although the orientation of d-lines is altered by the inclination of weak links, the observable angles are essentially the same.

Figure 4 exemplifies the evolution of the d-lines as a function of transparency.
Magneto-optical images in panels (a) to (d) were taken at temperatures ranging from 5.0~K to 7.5~K, using the sample with $\Phi = 65^\circ$. 
The transparency decreases from $0.35 \pm 0.03$ to $0.14 \pm 0.03$, as shown previously in the inset Fig. \ref{Fig3}(a).
Panels (e) to (h) depict the reconstruction of the d-lines and streamlines for values of $\tau$ representative of the interval which includes those obtained experimentally for $\Phi = 65^\circ$, namely, 0.35, 0.29, 0.21, and 0.14, respectively.

\begin{figure}
\centering\includegraphics[width=1.0\linewidth]{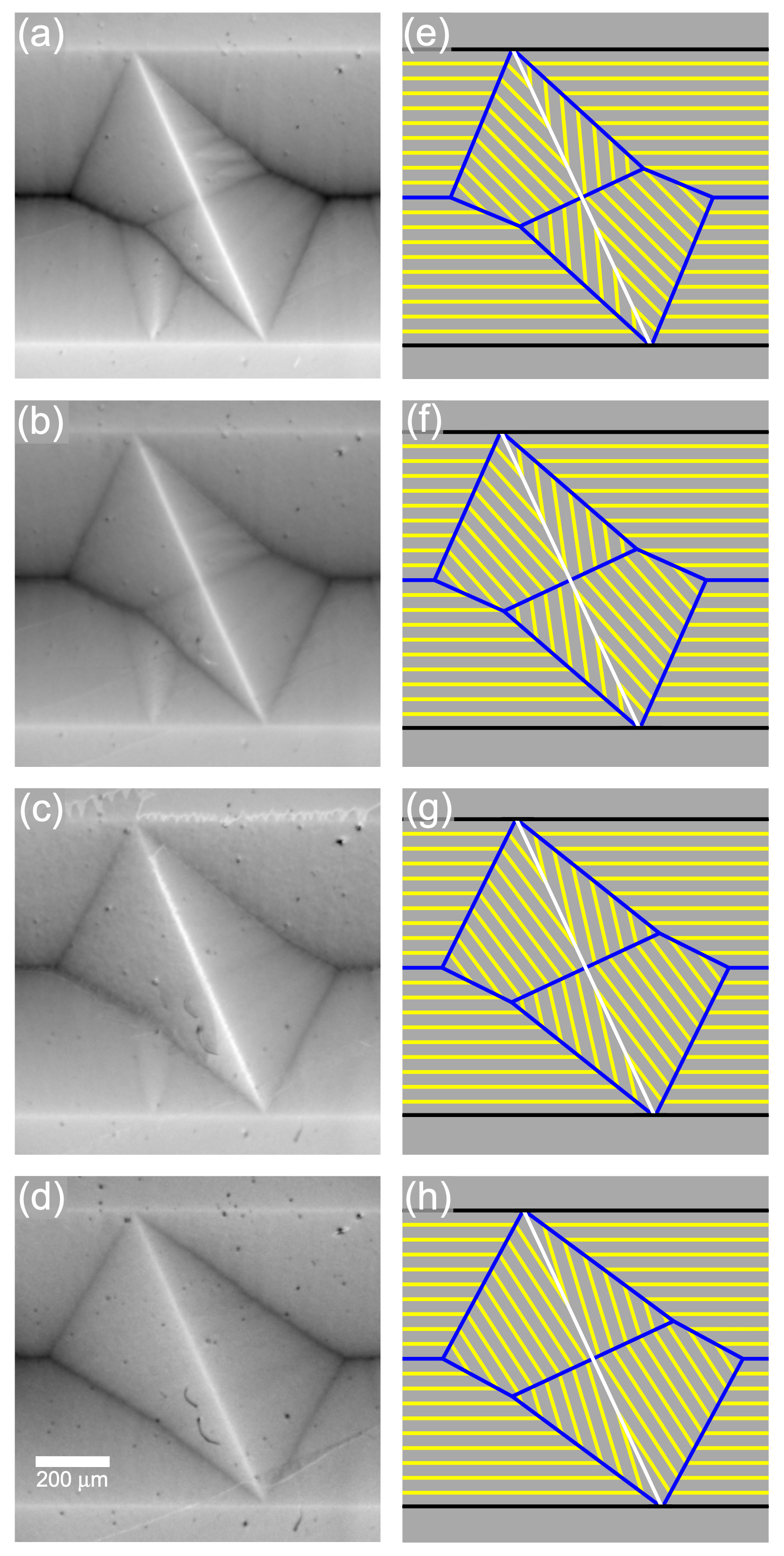}
\caption{ Left~column: Magneto-optical images of the film with $\Phi~=~65^\circ$ at: (a)~7.5~K and 40~Oe, (b)~7.0~K and 65~Oe, (c)~6.0~K and 145~Oe, and (d)~5.0~K and 200~Oe.
Right~column: Reconstruction of streamline patterns with corresponding transparencies: (e)~0.14, (f)~0.21, (g)~0.29, and (h)~0.35.}
\label{Fig4}
\end{figure}

As $\tau$ increases, $\theta$ decreases, causing the domains in the neighborhood of the WL to shrink. 
At lower $\tau$, the dark line crossing perpendicularly the WL at its midpoint is less intense. 
This faint line is also a d-line, where the bending angle of the streamlines tends to $0^\circ$. In fact, at higher temperatures, consequently, lower $\tau$, the bending angle of the current streamlines at this faint line becomes so minimal that the d-line disappears. 
Thus, a darker d-line indicates a more pronounced curvature of the streamlines.

\begin{figure}[tb!]
\centering\includegraphics[width=1.0\linewidth]{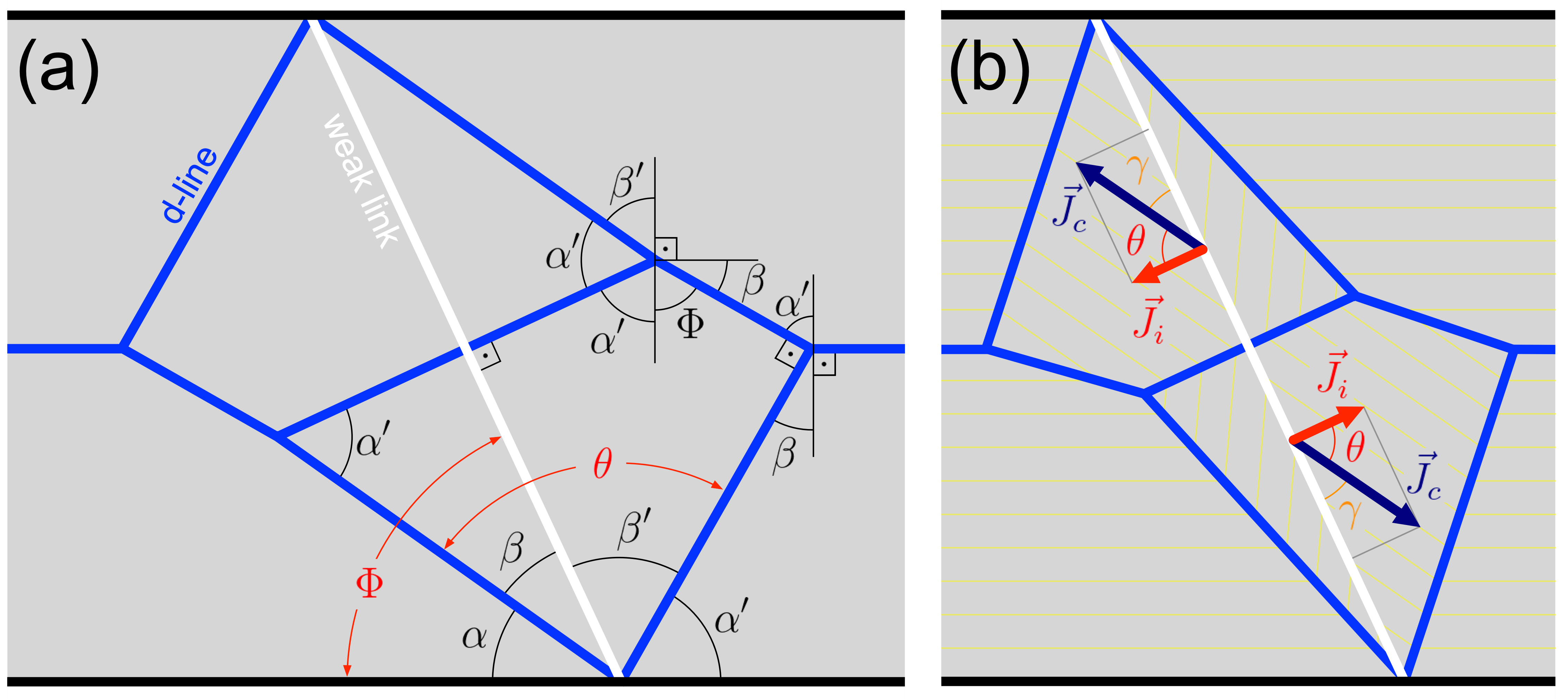}
\caption{(a) Schematics of WL region, showing angles and d-lines (blue). The internal angles can be defined using $\theta$ and $\Phi$, as per Eqs.~\ref{beta} to \ref{alpha-prime}. The drawing is top-bottom symmetric, so the angles on the upper part are omitted. (b) The angle formed by the critical currents $J_i$ and $J_c$ at the WL is $\theta$, what visually confirms Eq. \ref{Eq12}.}
\label{Fig5}
\end{figure}

All internal angles in Fig. \ref{Fig5}(a) can be expressed in terms of $\theta$ and $\Phi$, both of which are easily observable. The angle $\beta$ is
\begin{equation}
    \beta =\frac{1}{2}\left ( \theta +\Phi - 90^{\circ} \right ).
    \label{beta}
\end{equation}

Once $\beta$ is determined by the Eq.\ref{beta}, the other angles can be calculated as follows:
\begin{equation}
    \beta'=\beta -\Phi +90^{\circ},
    \label{beta-prime}
\end{equation}
\begin{equation}
    \alpha=\Phi-\beta,
    \label{alpha}
\end{equation}
and
\begin{equation}
    \alpha'=90^{\circ}-\beta.
    \label{alpha-prime}
\end{equation}

Therefore, measuring $\theta$ for a given $\Phi$ determines not only $\tau$, using Eq. \ref{Eq12}, but also, through Eqs. \ref{beta} to \ref{alpha-prime}, all d-lines related to the presence of the WL. 

To complete this detailed analysis of the WL neighborhood, we refer to Fig. \ref{Fig5}(b) to visually represent the true meaning of Eq. \ref{Eq12}, i.e., $\tau=cos\theta$: the WL and pristine critical currents, $J_i$ and $J_c$, respectively, are shown at two symmetrical points on the WL, enclosing the angle among them which, according to Fig. \ref{Fig2} and Eq. \ref{Eq011}, is $\theta$. The current density that crosses the WL is, by definition, perpendicular; hence we immediately see that Eq. \ref{Eq12} represents the relationship between transparency and the observable angle, regardless of the tilt angle.

According to Eq. \ref{Eq12}, a tilted weak link does not impact the relationship between transparency and $\theta$, since $\tau$ is independent of $\Phi$. 
However, this angulation restricts the maximum transparency what, according to Eqs. \ref{Eq12} and \ref{beta}, occurs for $\beta=0$:

\begin{equation}
    \tau _{max}=\displaystyle \lim_{\beta \to 0} \cos \left( 2\beta - \Phi  + 90^{\circ} \right )= \sin \Phi.
    \label{tau-max}
\end{equation}

It follows that, for angle-jointed samples, the upper limit for the current carried from grain to grain through the WL may be lower than the critical current of the WL, as the transparency cannot exceed sin~${\Phi}$. The physical meaning of this result can be readily grasped from a closer inspection of Fig. \ref{Fig2}: as $\tau_{max}$ requires $\beta=0$, triangle 2 vanishes in this condition, and $J_c$ arrives at the WL horizontally. Thus, current continuity implies that $J_i$ has to match exactly the $J_c$ projection normal to the interface, namely, $J_i = J_c \sin\Phi$, since the angle between the vertical direction and the normal to the WL is equal to that between the horizontal direction and the WL itself, i.e., $\Phi$. This finding has technological implications for joining bulk superconductors: the faces to be connected must be prepared in such a way to ensure  $\Phi = 90^\circ$ for optimal current flow. As a matter of fact, such a strong angular dependence of the critical current has been demonstrated previously in studies with misoriented bicrystals for which the tilt angle $\Phi$ is typically in the range from $70^{\circ}$ to $90^{\circ}$~\cite{Polyanskii_Magneto_1996,Guth2004,palau2004simultaneous,palau2007simultaneous}.

Minimum transparency occurs when the two sides are disconnected, $\tau=0$, resulting in an angle $\theta$ = $90^{\circ}$. 
Under these conditions, $\beta$ reaches its maximum value and is the bisector of the angle between the WL and the sample edge.

\begin{equation}
    \beta_{max} = \frac{1}{2} \Phi.
    \label{beta-max}
\end{equation}

Figure 6 represents this situation for the three values of the tilt angle studied here: $\Phi = 90^\circ, 65^\circ, 45^\circ$. The streamlines are colored yellow and red on each side of the WL to emphasize that there is no current leakage through the connection.

\begin{figure}[tb!]
\centering\includegraphics[width=0.9\linewidth]{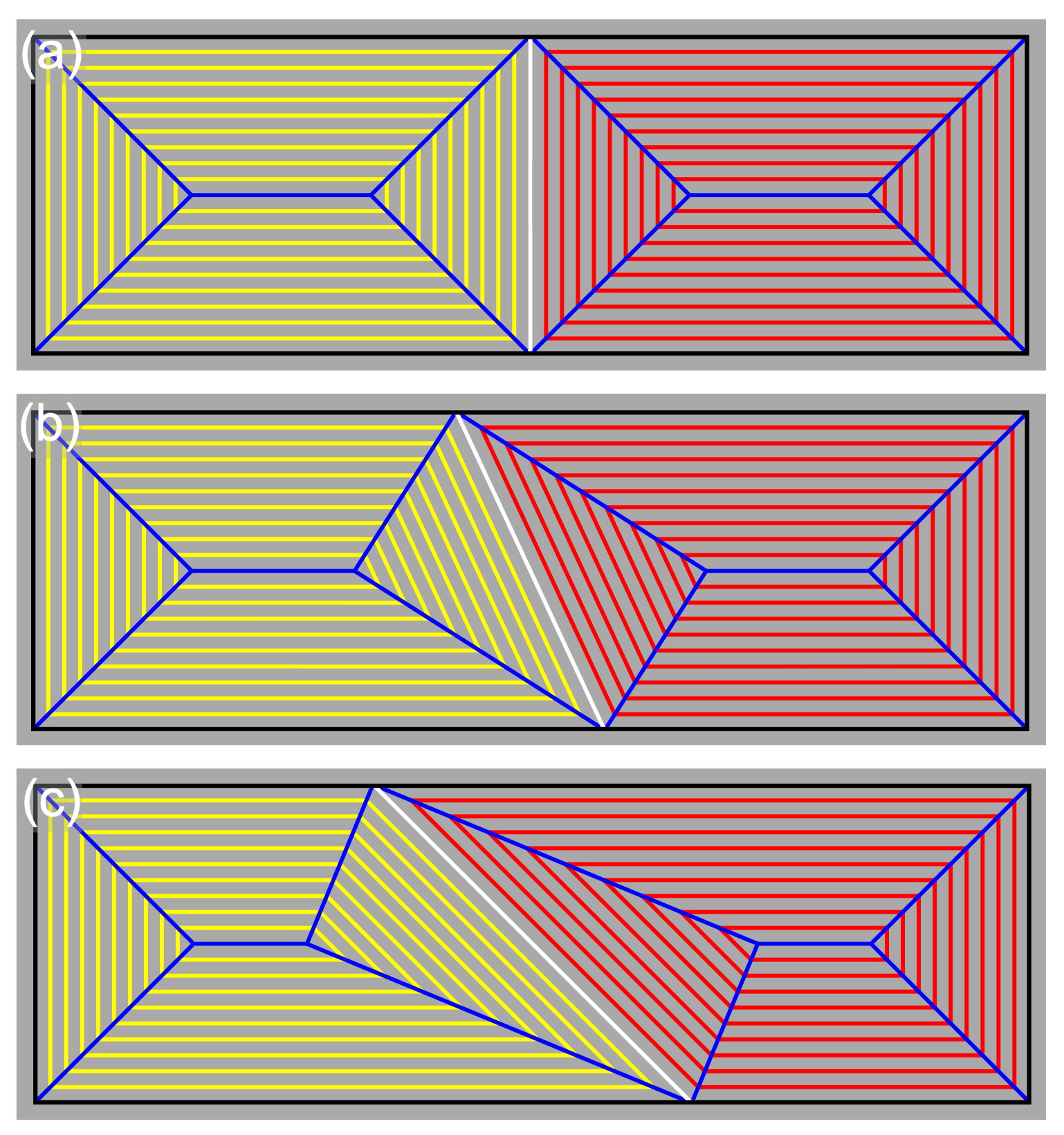}
\caption{Reconstruction of critical current streamlines for $\tau~=~0$.
The WLs are tilted by $\Phi$~= (a)~90$^{\circ}$, (b)~65${^\circ}$, and (c)~45$^\circ$.
Since the grains are disconnected, so are the left (yellow) and right (red) sets of streamlines.
}
\label{Fig6}
\end{figure}

The method described here for measuring the electrical connectivity in superconductors with WLs can also be applied to systems with four plates, thus creating a larger superconducting area. 
The V-shape formed by a pair of d-lines exhibits the same characteristics as those shown in Fig. (\ref{Fig1}) for a perpendicular WL. 
Eq. (\ref{Eq12}) remains valid~\cite{colauto_measurement_2021}, now generalized for WLs at oblique angles.

\section{Summary}

Using the Bean model, the d-lines are geometrically reconstructed by defining \(\tau\). 
The precision of the drawing is determined by the arbitrarily chosen spacing $c$. 
Once  $\tau$ and $c$ are set, we calculate the streamline spacing along the weak link as $i = c/\tau$ and the angle $\gamma=\sin^{-1}\tau$. 
The streamlines, which are apart by a distance $c$ far from the WL, keep this distance $-$ characteristic of the critical current of the grain $-$ in the domain areas, but have to adapt to a distance $i$ along the interface. In turn, $i$ and the angle $\gamma$ at the intersection lead to the spontaneous appearance of the d-lines, which accurately match those obtained from the magneto-optical images. 
Experimentally, transparency is determined by measuring the easily observable angle $\theta$, allowing for the calculation of all other angles related to the d-lines. 
The tilting of the WL does not affect the relationship between transparency and $\theta$, but does limit the maximum transparency of a junction to sin~$\Phi$. 
This effect can be harnessed in technological solutions to, for instance, intentionally limit the current by creating a beveled junction or to avoid current limitation by ensuring a right-angle interface.

\begin{acknowledgments}
Brazilian agencies S\~ao Paulo Research Foundation (FAPESP, grants 2016/12390-6, 2017/24786-4 and 2021/08071-8), Coordena\c{c}\~ao de Aperfei\c{c}oamento de Pessoal de N\'ivel Superior - Brasil (CAPES) - Finance Code 001, and National Council of Scientific and Technologic Development (CNPq, grants~302564/2018-7, 302586/2018-0, 130831/2018-2, 309928/2018-4, 306894/2022-0, 316602/2021-3, 431974/2018-7, and 434797/2018-9).
The authors are grateful to T. H. Johansen for inspiring discussions on the MOI technique and also on the physics content of the paper.
\end{acknowledgments}

\bibliography{apssamp}

\end{document}